\documentclass{sig-alternate-05-2015}

\usepackage{hyperref}
\begin{document}

\setcopyright{acmcopyright}

\conferenceinfo{Web Science '16}{}

%


\title{User fluctuation in communities: a forum case}
%
%
%
%
%

\numberofauthors{1} 
%
\author{
%
%
\alignauthor
Zinayida Petrushyna\\
       \affaddr{RWTH Aachen University}\\
       \affaddr{Ahornstrasse 55}\\
       \affaddr{52056, Aachen, Germany}\\
       \email{petrushyna@dbis.rwth-aachen.de}
}


\maketitle
\begin{abstract}
Understanding fluctuation of users help stakeholders to provide a better support to communities. Below we present an experiment where we detect communities, their evolution and based on the data characterize users that stay, leave or join a community. Using a resulted feature set and logistic regression we operate with models of users that are joining and users that are staying in a community. In the related work we emphasize a number of features we will include in our future experiments to enhance train accuracy. This work represents a first from a series of experiments devoted to user fluctuation in communities.
\end{abstract}

%
%


%
%

%
%



\section{Introduction}
Popular discussions in social media are engaged by most of users that are not directly devoted to discussed topics. Anyway Internet citizens are not stuck on one community and its discussion topic only and usually move rapidly from one community to the other. at someday popular discussions about the Russian-Ukrainian conflict do not bother masses anymore while audience move to other forum pages to discuss refugees problems. What is the reason of such a change? and how can we foreseen user fluctuation from one community to the other. Political topics are definitely event-driven and highly depend on mass media \cite{CPHa12}. In our experiments we are interested more in user fluctuation in communities devoted to odd topics such as hobbies, jobs or life long learning. To answer this question we investigate a forum of life long learners in medicine domain. Further we explain our experimental setup, our first results and work related to our experiment and contributing to our future work. 

\section{Experiment Setup and Results}

In this experiment we focus on forums of the Student Doctor Network\footnote{Student Doctor Network \url{http://forums.studentdoctor.net/}} that are dedicated to all medical students and personnel in the life long learning. The collected in the Mediabase \cite{KlPe08} data include posts for the time period of 10 years, from April, 21 2000 till April 25, 2013. We operate with 208K posts, 25K users, 8K threads where the average depth of a thread is 25 posts. 

The purpose of models we create in the experiment is to differ between users that are going to join a community (\textit{joining} users), users that are in a community where others are going to join (\textit{previous} users), users that are going to leave a community (\textit{leaving} users) and users that are staying in a community although others are leaving (\textit{staying} users). Using such models we can predict behaviors of current users and anticipate it. 

Our fist step in the experiment is to detect communities using a propinquity algorithm \cite{ZWWZ09}, their evolution detecting evolved states according to \cite{APUc09} and based on these define joining, previous, leaving and staying users.

We divide our data set into 199 snapshots that include posts from the time interval of 24 days. Based on these posts we define connections between users if they participate in the same thread. We detected 1286 communities in the given period where the highest number of communities in a snapshot is 33 while 69 snapshots have no communities defined. All detected communities include 31K users where 16K are unique users. 

Based on \cite{APUc09} we differentiate between users considering their attitude to communities. \textit{Joining users} 1)are joining a community at current time, 2) have not appeared in the community before (in the previous snapshot) and 3)the community includes more than 50\% of nodes in the current snapshot that have been in the community in the previous snapshot. The nodes that have been staying all the time in the community are called \textit{previous users}. \textit{Leaving users} have participated in a community in a previous snapshot but do not appear in the community in a current snapshot. While other, \textit{staying}, users of the community (at least 50\%) appear both in the previous and current snapshot.

In the previous study \cite{PKKr15} we investigate all users and communities in StDoctorNet forum and count their connectiveness and betweenness \cite{WaFa94}. Furthermore, we consider user posts for calculating their sentiments and their attitude to knowledge. For this purpose we imply LIWC dictionary \cite{PCI*07} with the help of which we detect words indicating frustration, anger, satisfaction and cognitive work and  create language models that help to detect how positive, negative or cognitive a post is. Moreover, one further feature was extracted while mining user posts. We detect phrases of users that indicate their intentions from a linguistic point of view \cite{KPKK11,StKr12}. 


Other features that were added in the current feature set characterize user behavior in a community earlier, before the current state. Furthermore, we added modularity \cite{NeGi04} of a snapshot where a user appears to the set as well. The list of features can be found in the Table\~{tab:features}.
\begin{table}
\centering
\caption{A list of features}
\begin{tabular}{|p{3cm}|p{5cm}|} \hline
Feature name&Description\\\hline
sentiment&number of words denoting sentiments according to the LIWC dictionary\\\hline
cognition&number of words devoted to a cognitive work according to the LIWC dictionary\\\hline
intent&the number of intent phrases a user posted\\\hline
connectiveness&the network measure that shows how close/far a node to the center is\\\hline
betweenness&the network measure indicating frequency of node appearance on short paths\\\hline
number of times appeared before& the number of posts a user wrote before the current time\\\hline
avgSentimentBefore&average sentiment measure of a user that is based on her posts published before the current time\\\hline
avgCognitionBefore&average cognition measure of a user that is based on her posts published before the current time\\\hline
avgIntentBefore&average intent measure of a user that is based on her posts published before the current time\\\hline
avgConnectiveness&average connectiveness measure of a user from snapshots before the current time\\\hline
avgBetweenness&average connectiveness measure of a user from snapshots before the current time\\\hline
lastSentiment&last sentiment measure before the current time\\\hline
lastCognition&last cognition measure before the current time\\\hline
lastIntent&last intent measure before the current time\\\hline
lastConnectiveness&last connectiveness measure before the current time\\\hline
lastBetweenness&last betweenness measure before the current time\\\hline
last activity& the number of days left after a user posted his last post before the current time\\\hline
modularity&the measure that shows the tightness of all links of a snapshot where a user appears\\
\hline\end{tabular}
\label{tab:features}
\end{table}

We create logistic regression models using the given feature set. First of all, the data was normalized. After that it was divided into train and cross-validation sets using Monte Carlo cross-validation. We present our results listing precision, recall and F-measure for models with different features in Table~\ref{tab:models}. Values of precision, recall and F-measure are promising, though train accuracy for any model is close to 50\% that indicates the requirement to enhance the feature set and the number of train examples as well as to brush the data to avoid outliers (e.g., users with high betweenness). 

\begin{table}
\centering
\caption{A list of applied models}
\begin{tabular}{|p{3cm}|p{1.5cm}|p{1.5cm}|p{1.5cm}|} \hline
Model&Precision&Recall&F-measure\\\hline
M1: all features&\textbf{0.6870}&0.5716&0.6240\\\hline
M2: M1 without current sentiment, cognition, intent&0.6489&0.6681&0.6584\\\hline
M3: M1 without any sentiment, cognition, intent&\textbf{0.6984}&0.6260&\textbf{0.6602}\\\hline
M1 without modularity&0.5888&0.6462&0.6162\\\hline
M3 without avgconnectiveness and avgbetweenness&0.6021&0.6862&\textbf{0.6414}\\
\hline\end{tabular}
\label{tab:models}
\end{table}

\section{Related  and Future Work}

In the following section we observe related works and features from recent literature. The features are potential candidates to be added into our feature set. 

\cite{AlLe15} analyze donor communities and use standard machine learning techniques to predict donor return. We can replicate some of the features the scholars are using to enhance our feature set since any forum user is donating by sharing knowledge. \textit{Donation} of a forum user is an amount of posts that appear in the past in a forum. Acceptance of these posts or, in other words, attitude of a community can be graded by a response rate. The time used by the scholars to predict the return can be useful in our case since some of communities in StDoctorNet are devoted to exams that appear in a particular period of time. 

Since we pursuit to create models using both structural and semantic measures, experiences of \cite{ChCh15} can be as well considered. A bag-of-words is one of the collection of features that include only frequent words and help to classify spam messages but can be helpful to classify other cases as well. Furthermore, number of hyperlinks, quotations of other users, emoticons, punctuation characters, number of lines and words in a post - all these play a role for the spam detection and can be replicated in our experiment. 

Predicting replying behaviors is a challenging question in any kind of social media. \cite{KAG*15} investigate e-mail networks and a variety of factors that affect reply time and length. Such factors like circadian rhythms, demographics of users, posting from mobile devices infer different behaviors of users and therefore enlarge their probability in joining a community. Furthermore, the authors find the evidence of the synchronization in dyadic interactions within a thread that is worse to investigate further in the scope of a forum.

We can not skip the discussion of the \textit{link prediction problem} related to the user fluctuation. The problem solution is proposed in numerous approaches that foresee links that will appear in the future and connect unconnected nodes or group of nodes. These links can then define users that are joining or leaving communities. In \cite{LeHK10b} authors focus on sign networks that predict links operating with positive or negative connections between nodes. One of possible feature can be a number of positive and negative links a user has or a number of positive and negative links a friend of a user (a connected node) has. The latter emphasizes a further set of features devoted to neighborhood. The features may include information about previous connections of users, their positive or negative attitude, neighborhood fidelity to a forum, etc.

Further perspective discussed in the literature is the influence of first impression of a user on her future cooperation in a forum. Therefore, \cite{DiFa15} investigate self-disclosure patterns in forums and their influence on further participation of users in the forums. Trust is tremendously important for users in community participation. Absence of trust is one of the reasons of overwhelming lurking. The measurement of the first impression or experience of a user together with trust to a community can be a good addition to our future set of features. 

In this paper we have presented only a preliminary work on prediction of users that are leaving or joining communities. We list our features and enumarate a number of features that can be used to enlarge the feature set. The results of classifiers can be used for forum communities to detect problems in community populations and manage discussionsS.

\bibliographystyle{abbrv}
\bibliography{fluctuation}  
\end{document}